\documentclass[%
 aip,
 amsmath,amssymb,
 reprint,%
]{revtex4-1}

\newcommand{\IMSS}{Muon Science Laboratory, Institute of Materials Structure Science, High Energy Accelerator Research Organization (KEK-IMSS), Tsukuba, Ibaraki 305-0801, Japan}
\newcommand{\Sokendai}{Graduate Institute for Advanced Studies, SOKENDAI}
\newcommand{\Ibadai}{Graduate School of Science and Engineering, Ibaraki University, Bunkyo, Mito, Ibaraki 310-8512, Japan.}
\newcommand{\Kinken}{Institute for Materials Research, Tohoku University (IMR), Katahira, Aoba-ku, Sendai 980-8577, Japan}
\newcommand{\Kyudai}{Department of Applied Chemistry and Center for Polymer Interface and Molecular Adhesion Science, Kyushu University, Fukuoka 819-0395, Japan}

\usepackage{multirow}
\usepackage{graphicx}
\usepackage[dvips]{color}
\usepackage{dcolumn}
\usepackage[version=3]{mhchem}
\usepackage{txfonts}
\usepackage{bm}
\usepackage{ulem}

\usepackage[utf8]{inputenc}
\usepackage[T1]{fontenc}
\usepackage{mathptmx}
\usepackage{etoolbox}
\usepackage[hypertex,colorlinks=true,linkcolor=black,citecolor=blue,filecolor=blue,urlcolor=blue,setpagesize=false,nesting=true]{hyperref}

\makeatletter
\def\@email#1#2{%
 \endgroup
 \patchcmd{\titleblock@produce}
  {\frontmatter@RRAPformat}
  {\frontmatter@RRAPformat{\produce@RRAP{*#1\href{mailto:#2}{#2}}}\frontmatter@RRAPformat}
  {}{}
}%
\makeatother

\begin{document}
\title{Slow polymer dynamics in poly(3-hexylthiophene) probed by muon spin relaxation}

\author{S.~Takeshita}\affiliation{\IMSS}\affiliation{\Sokendai}
\author{K.~Hori}\thanks{Present Address: Sumitomo Rubber Industries Ltd., Kobe, Hyogo 651-0072 Japan}\affiliation{\IMSS}
\author{M.~Hiraishi}\affiliation{\IMSS}\affiliation{\Ibadai}
\author{H.~Okabe}\affiliation{\IMSS}\affiliation{\Kinken}
\author{A.~Koda}\affiliation{\IMSS}\affiliation{\Sokendai}
\author{D.~Kawaguchi}\thanks{Present Address: Department of Chemistry and Biotechnology, Graduate School of Engineering, The University of Tokyo, Bunkyo-ku,
Tokyo 113-8656, Japan}\affiliation{\Kyudai}
\author{K.~Tanaka}\affiliation{\Kyudai}
\author{R.~Kadono}\thanks{email: ryosuke.kadono@kek.jp}\affiliation{\IMSS}
\date{\today}

\begin{abstract}
The molecular dynamics of regioregulated poly(3-hexylthiophene) P3HT is investigated using muon spin relaxation ($\mu$SR). The response of the $\mu$SR spectra to a longitudinal magnetic field ($B_{\rm LF}$, parallel to the initial muon spin direction) indicates that the implanted muons form both muonated radicals localized on the thiophene ring and diamagnetic states with comparable yields. Moreover, the unpaired electron in the radical undergoes hyperfine interactions with muon bound to thiophene and with neighboring protons, whose fluctuations can serve as a measure for the molecular dynamics. The $B_{\rm LF}$ dependence of the longitudinal muon spin relaxation rate ($1/T_{1\mu}$) measured in detail at several temperatures is found to be well reproduced by the spectral density function $J(\omega)$ derived from the local susceptibility that incorporates the Havriliak-Negami (H-N) function used in the analysis of dielectric relaxation, $\chi(\omega)\propto1/[1-i(\omega/\tilde{\nu})^\delta]^\gamma$ (where $\tilde{\nu}$ is the mean fluctuation rate, and $0<\gamma, \delta\le1$). The magnitude of $\tilde{\nu}$ and its temperature dependence deduced from the analysis of $1/T_{1\mu}$ are found to be consistent with the motion of hexyl chains and thiophene rings suggested by $^{13}$C-NMR. The present result marks a methodological milestone in the application of $\mu$SR to the dynamics of complex systems with coexisting fluctuations over a wide range of time scales, such as polymers.
\end{abstract}
\maketitle

\section{Introduction}
Conjugated polymers have been widely studied as promising semiconductor materials because of their  potential in electronic devices such as solar cells \cite{Gunes:07,Thompson:08,Kim:06,Reynords:19}, field-effect transistors \cite{Dimitrakopoulos:02,Braga:09,Yan:09}, light-emitting diodes \cite{Burroughes:90,Krummacher:06}, and those for spintronics \cite{Naber:07,Vardeny:10,Geng:16}. In particular, their light weight, flexibility, and elasticity have attracted attention as materials for wearable devices unsuitable for inorganic semiconductors \cite{Kuribara:12,Kaltenbrunner:13}. Among them, regio-regular poly(3-hexylthiophene) (P3HT), in which the hexyl side chain is regularly bound from head to tail at the 3-position of the thiophene ring, has attracted much attention as an optoelectronic material due to its excellent properties and solubility in organic solvents \cite{Chen:95,Bao:96,Sirringhaus:99}.
Of particular importance for device applications are thin films with thicknesses on the order of 10--100 nm. However, the relationship between the thin film morphology of organic semiconductors and their electrical and optical properties is still largely unresolved due to the lack of suitable experimental methods for proper investigation.

The molecular organization of conducting polymers in the condensed state and their dynamical structure have a marked influence on their optoelectronic and electrical properties. For example, thermally induced disorder in the alkyl side groups of conjugated oligomers and polymers correlates with optical absorption of bithiazole oligomers \cite{Curtis:07}, which may be relevant with poly(3-alkylthiophene) (P3AT) with longer alkyl side-chains \cite{Bolognesi:93,Zhao:97}. While the melting behavior of the hexyl side group and related polymorphism has not been reported for P3HT,  thermochromism phenomena suggests that thermally activated thiophene twists from a quasi-ordered structure causes disorder in the $\pi$ conjugation \cite{Inganas:88}. Despite the importance of information on the local molecular dynamics, such information on P3HT is so far limited.

In this respect, recent Fourier transform infrared spectroscopy (FT-IR) and $^{13}$C nuclear magnetic resonance (NMR) studies report interesting results \cite{Yazawa:10}. They suggested that alkyl side chain motions are activated  in P3HT crystal above $\sim$200 K, and that twisting motions of the thiophene ring sets in above $\sim$300 K to exhibit a plastic crystal state. Similar indications are obtained from dynamic mechanical analysis which allows direct access to the thermo-molecular motion of the polymer in the film, where three relaxation peaks at around 200, 300, and 390 K (corresponding to the so-called $\beta$, $\alpha_1$, and $\alpha_2$ processes) are observed in the dynamic loss modulus \cite{Ogata:15}.  From the magnitude of the apparent activation energy and from the complementary structural analysis using FT-IR, these processes are assigned to side chain motion, twisting motion, and deformation of inter-lamellar crystal, respectively.

Previous theoretical studies predict that the vibrational slow twisting and fast coupling stretching motions of $\pi$-conjugated molecules are strongly coupled to the electronic structure and are therefore related to photoelectric processes \cite{Tretiak:02}. In particular, twisting motion is also important in that it is thought to govern the efficiency of the generation of isolated polarons from photoinduced electron-hole polaron pairs that we have proposed \cite{Ogata:15}. Therefore, the details of the dynamical structure of P3HT and the relationship between its structure and physical properties such as optical absorption remain one of the most important issues.

Here, we report on local molecular dynamics inferred from muon spin relaxation ($\mu$SR) experiments in the crystalline state P3HT. In general, the positive muon ($\mu^+$) implanted into a material behaves as a light isotope of hydrogen (denoted hereafter by the element symbol Mu). We show from the response of $\mu$SR spectra to the longitudinal magnetic field (LF, whose magnitude given by $B_{\rm LF}$) that about a half of implanted muons form a muonated radical state and that the unpaired electron in the radical exerts hyperfine fields to both $\mu^+$ and surrounding protons which are described by the hyperfine (HF) and nuclear hyperfine (NHF) parameters.  The unpaired electron can also be viewed as a state in which charge carriers doped by muon addition reactions (equivalent to hydrogenation) are localized in the vicinity of muons. In terms of defects in solids, such a state is approximately equivalent to the polaron often referred to in the field of conducting polymers \cite{Mizes:93,Furukawa:23}.  $\mu$SR has the advantage of providing detailed information on the local electronic structure and dynamic properties of such polaron states produced by muons themselves.  We also show that the remaining diamagnetic muons show negligible spin relaxation.


The time-dependent fluctuation of the HF and/or NHF fields, which causes longitudinal muon spin relaxation, can serve as a measure of local molecular dynamics at the muon site(s). We show that the LF dependence of the spin relaxation rate, $1/T_{1\mu}$,  is reproduced by the spectral density function $J(\omega)$ derived from the local susceptibility that incorporates the Havriliak-Negami (H-N) function, $\chi(\omega)\propto1/[1-i(\omega/\tilde{\nu})^\delta]^\gamma$  (where $\tilde{\nu}$ is the mean fluctuation rate, $0<\gamma, \delta\le1$), used in the analysis of dielectric relaxation \cite{Havriliak:67,Alvarez:91}. The qualitative change in the LF dependence of $1/T_{1\mu}$ observed around 200 K is typically represented by the behavior of  $\tilde{\nu}$, and its temperature dependence are found to be consistent with those due to the molecular motions suggested by previous study using $^{13}$C-NMR, where the characteristic time scale of the fluctuation ($\tilde{\nu}^{-1}$)  is deduced to be $10^{-9}$--$10^{-10}$~s above $\sim$200 K. 

We also find that the amplitude of the internal field fluctuations is significantly smaller than the linewidth $\Delta_{\rm n}$ of the NHF (or HF) field at lower temperatures. This corresponds to the expected situation that the muonated radicals are stationary and only a part of the host molecules are in motion, suggesting that the strong collision model (random phase approximation) commonly assumed in dynamical models does not hold. We show that the recently proposed model to distinguish ion dynamics from muon self-diffusion under similar circumstances \cite{Ito:24} can be extended to paramagnetic Mu to provide reasonable accounts of the experimental results on P3HT.

The present result also suggests a need to reconsider the earlier interpretation that $1/T_{1\mu}$ in P3HT is due to the fluctuation of HF fields induced by rapid quasi-one-dimensional jumping of carriers (unpaired electrons) created on the thiophene chain upon muon implantation \cite{Risdiana:10a,Risdiana:10b}.  

 \section{EXPERIMENTAL}

\par
The regioregular P3HT samples were prepared using the compound purchased from Sigma-Alidrich Inc.~as received. Number-average molecular weight ($M_{\rm n}$) and polydispersity index (PDI) were 26  kg mol$^{-1}$ and 2.4, respectively, which were determined by gel permeation chromatography using polystyrene standards. The melting temperature measured by differential scanning calorimetry was 488 K. Films of P3HT were prepared by spin-coating from chloroform solutions onto quartz, barium fluoride (BaF$_2$) and commercially available polyimide (PI)
substrates for absorption spectroscopy, FT-IR measurements and dynamic
mechanical analysis, whose results were reported elsewhere \cite{Ogata:15}. The films were dried under vacuum at 373 K for 12 h. The film thickness was approximately 300 nm.

Conventional $\mu$SR measurements were performed of the P3HT films (25 mm$\phi\times0.7$ mm) using the S1 instrument (ARTEMIS) at the Materials and Life-science Experiment Facility, J-PARC  \cite{ARTEMIS}, where high-precision measurements over a long time range of 20 $\mu$s can be routinely performed using a high-flux pulsed beam of positive muons ($\approx10^3$--$10^4$ $\mu^+$s per pulse, with a repetition rate 25 Hz and an incident energy $E_\mu\approx4$ MeV).
The $\mu$SR spectra [the time-dependent decay-positron asymmetry, $A(t)$] which reflects the magnetic field distribution at the Mu site, was measured from 100 K to 350 K under zero field (ZF), weak LF (parallel to the initial Mu polarization ${\bm P}_\mu$), and weak transverse field (TF, perpendicular to ${\bm P}_\mu$), and were analyzed by least-squares curve fitting~\cite{musrfit}. The correction factor for the instrumental asymmetry under a given LF was calibrated by a separate set of LF-$\mu$SR measurements on a blank sample holder (made of silver), combined with the total asymmetry corresponding to 100\% muon spin polarization determined by a weak TF-$\mu$SR measurement.

\section{RESULT}
In conjugated polymers, implanted Mu can cleave double bonds to form a bound state with cation atom: e.g., for polyacetylene, 
\begin{flushleft}
\hspace{1em}\ce{[HC=CH]$_n$ + Mu} $\rightarrow$ \\
\hspace{4em}\ce{[HC=CH]$_{k}$-[CHMu-$\dot{\rm {C}}$H]-[HC=CH]$_{n-k-1}$},\\
\end{flushleft}
where ``$\cdot$'' on C denotes the unpaired electron, whose polaron motion has been studied by $\mu$SR in a variety of conducting polymers \cite{Nagamine:84,Ishida:85,Pratt:97,Blundell:02,Pratt:04}. Here, the Mu forms a covalent bond with carbon to form a closed electron shell (i.e., the ``diamagnetic Mu state''). However, when the unpaired electron is localized at adjacent atoms as shown above, Mu undergoes relatively large HF interaction with the unpaired electron, comprising a paramagnetic defect called a ``muonated radical'' (denoted Mu$\dot{R}$).

In general, the spin Hamiltonian for Mu$\dot{R}$ under an external magnetic field $B$ (parallel with $z$) is written 
\begin{equation}
\mathcal{H}/\hbar = 
\frac{1}{4}\omega_0{\bm \sigma}\cdot{\bf \tau} +\frac{1}{4}\omega_*({\bm \sigma}\cdot{\bm n})({\bm \tau}\cdot{\bm n}) -\frac{1}{2}\omega_\mu\sigma_z + \frac{1}{2}\omega_e\tau_z\nonumber 
\end{equation}\vspace{-6mm}
\begin{equation}
+\sum_m[\Omega_{\perp_m}{\bm S}_e\cdot{\bm I}_m +(\Omega_{\parallel_m}-\Omega_{\perp_m})S_e^zI_m^z]- \sum_m\omega_{n_m}I_m^z,\label{hn}
\end{equation}
where $\omega_0$ and $\omega_*$ respectively denote the transverse ($\omega_\perp$) and anisotropic ($\omega_\parallel-\omega_\perp$) parts of the HF interaction, $\omega_\mu=\gamma_\mu B$ with $\gamma_\mu=2\pi\times135.53$ MHz/T being the muon gyromagnetic ratio, $\omega_e=\gamma_e B$ with $\gamma_e=2\pi\times28024.21$ MHz/T being the electron gyromagnetic ratio, ${\bm \sigma}$ and ${\bm \tau}$ are the Pauli spin operators for muon and electron, and ${\bm n}$ is a unit vector along the symmetry axis of the Mu$\dot{R}$ state \cite{Percival:79,Patterson:88}. $\Omega_{\perp_m}$ and $\Omega_{\parallel_m}-\Omega_{\perp_m}$ are the transverse and anisotropic parts of the NHF parameter for the $m$th nuclear spin $I_m$ ($=\frac{1}{2}$ for $^1$H), $\omega_{n_m}=\gamma_{n_m}B$ with $\gamma_{n_m}$ being the gyromagnetic ratio for the corresponding nuclei; $\Omega_{\parallel_m}$ and $\Omega_{\perp_m}$ are usually dominated by the Fermi contact interaction with an order of magnitude smaller contributions from magnetic dipolar interaction.

\begin{figure}[t]
  \centering
	\includegraphics[width=0.9\linewidth,clip]{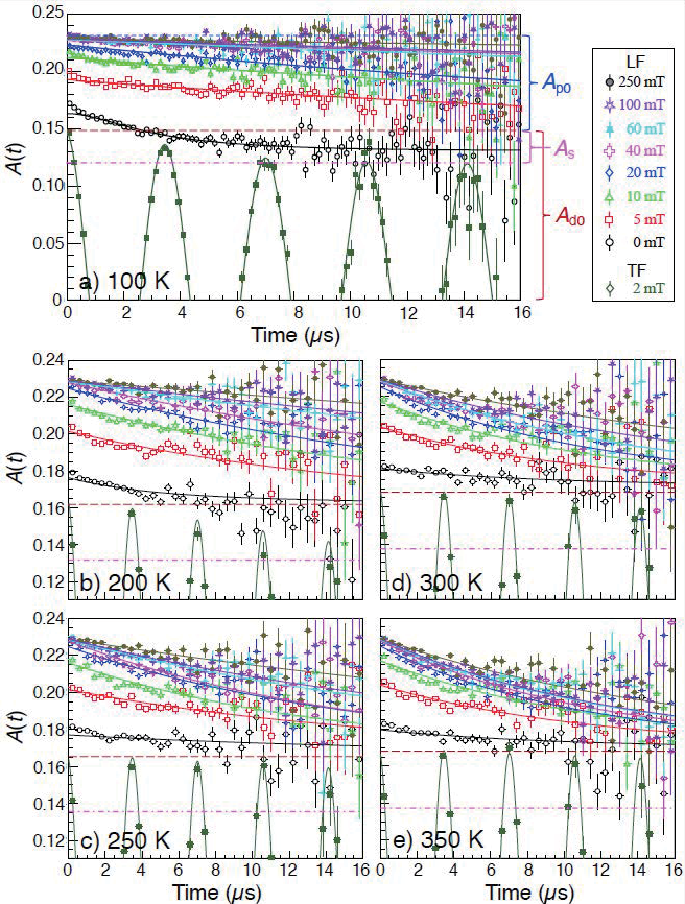}
	\caption{
	(a) TF- and ZF/LF-$\mu$SR time spectra observed at typical temperatures, which consists of paramagnetic ($A_{\rm p0}$) and diamagnetic ($A_{\rm d0}$) components, where the latter involves contribution from copper thermal shield showing relaxation ($A_{\rm s}$). The solid curves represent the least-square fits by Eqs.~(\ref{Atf}) and (\ref{Alf}). The horizontal dashed/dot-dashed lines show $A_{\rm d}$ and $A_{\rm s}$ determined by TF/ZF-$\mu$SR measurements. }
	\label{tspec}
\end{figure}

Figure \ref{tspec}(a) shows typical $\mu$SR time spectra under ZF, LF ($B_{\rm LF}=5$--400 mT), and TF ($B_{\rm TF}=2$ mT) [the portion for $A(t)>0$] observed at 100 K. These spectra consist of two components: the one that shows the Larmor precession under $B_{\rm TF}$ with a small fraction of exponential damping and another component with gradual recovery of $A(0)$ with increasing $B_{\rm LF}$.  The former includes background due to muons stopped in the materials around the sample, and its partial asymmetry ($A_{\rm b}$) is estimated to be $\approx0.067$ from calibration measurements. The initial asymmetry corresponding to 100\% spin polarization was $A_0\simeq0.23$ on the blank sample holder.

In order to use Mu as a probe, it is often crucial to know the electronic state of Mu in the target material. To this end, $\mu$SR measurements under a weak transverse field ($B_{\rm TF}$) is useful to evaluate the yield of the diamagnetic Mu state (Mu$^+$ or Mu$^-$ for the isolated Mu) by the initial asymmetry distribution in the $\mu$SR time spectra:
\begin{equation}
A(t)\simeq A_{\rm p}\sum_{i,j,k} f_{ijk}\cos\omega'_{ijk}t +A_{\rm d0}e^{-\lambda t}\cos\omega_\mu t,
\end{equation}
where $A_{\rm p}$ [$\equiv A(0)-A_{\rm d0}$] is the partial asymmetry of paramagnetic component (= Mu$\dot{R}$), $A_{\rm d0}$ is that of diamagnetic component with the damping rate $\lambda$, $\omega'_{ijk}$ and $f_{ijk}$ are the transition frequencies between the relevant Mu energy levels and their relative amplitudes ($i,j=1$--4 for the HF levels, $k$ is the number of sub-levels due to the NHF interaction which depends on ${\bm I}_m$) \cite{Patterson:88}. In the present measurements, the depolarization induced by the broad distribution of $\omega'_{ijk}$ due to the NHF interaction far exceeds the time resolution determined by the muon pulse width ($\tau\approx$80 ns at J-PARC, yielding the corresponding Nyquist frequency $1/2\tau\approx6.3$ MHz), and the first term is averaged out to yield
\begin{equation}
A(t)\simeq A_{\rm d0}e^{-\lambda t}\cos\omega_\mu t.\label{Atf}
\end{equation}

\begin{figure*}[t]
  \centering
	\includegraphics[width=0.8\textwidth,clip]{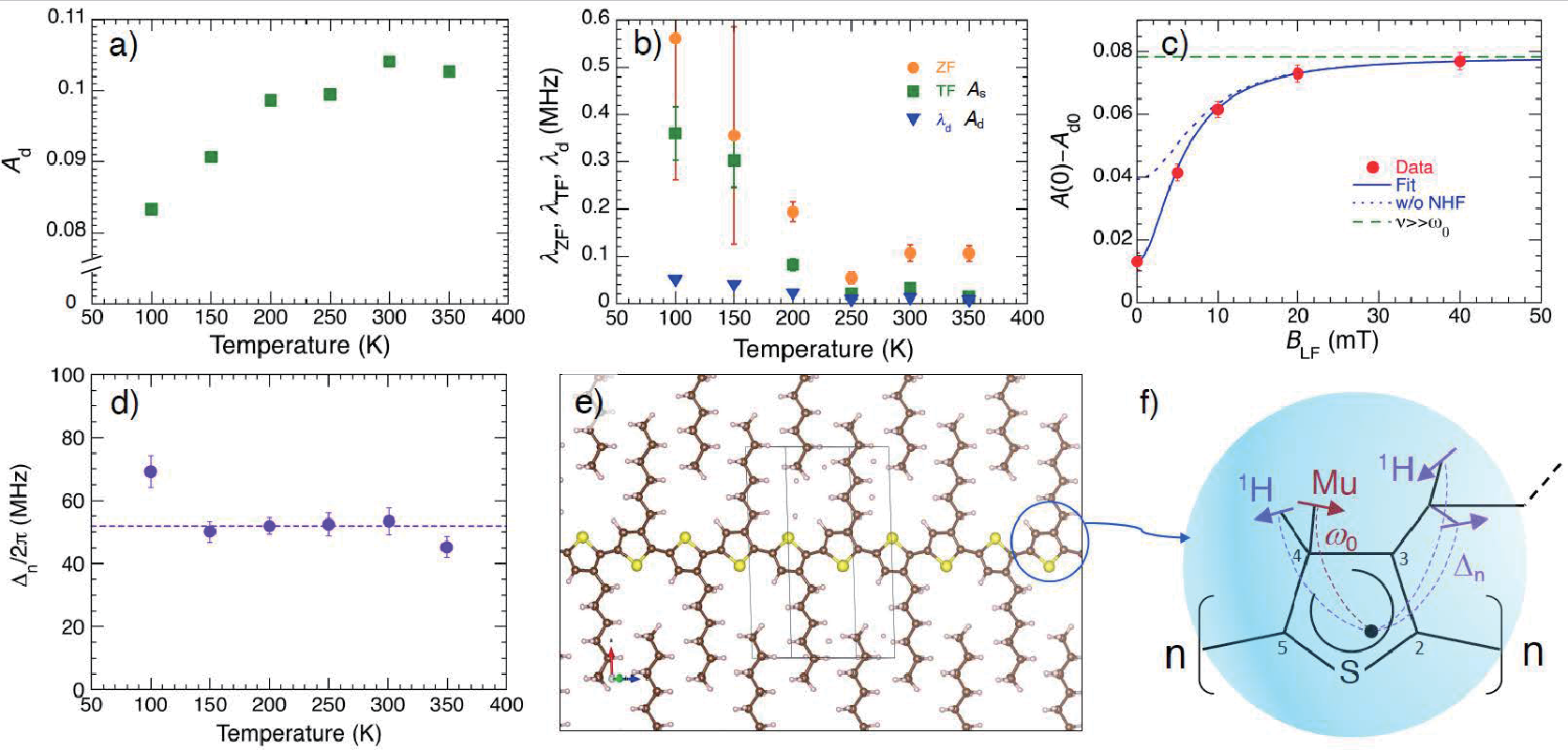}
	\caption{(a)--(b) Temperature dependence of the parameters for the diamagnetic Mu in P3HT: (a) partial asymmetry ($A_{\rm d}$), (b) transverse relaxation rate ($\lambda_{\rm TF}$) for the $A_{\rm s}$ component, and the upper bound of relaxation rate ($\lambda_{\rm d}$). The relaxation rate under a zero field ($\lambda_{\rm ZF}=1/T_{\rm 1\mu}$ in Eq.~(\ref{Alf}) with $B_{\rm LF}=0$) is also shown in (b) for comparison.  (c) An example for the initial asymmetry versus $B_{\rm LF}$ for the paramagnetic component obtained at 100 K, where solid line shows the result of fitting using Eq.~(\ref{gzp}), dashed line shows the case without NHF interaction, and dot-dashed line is for the case of fast fluctuations ($\nu\gg\omega_0$). (d) Temperature dependence of $\Delta _{\rm n}$ obtained by curve fitting. (e) Crystal structure of P3HT with yellow, brown, and white balls representing S, C, and H (after Ref.\cite{Yao:21}), and (f) a schematic illustration of the local electronic structure for Mu$\dot{R}$ and associated NHF interactions with nearby protons (arrows indicate their nuclear spins).}\label{hfp}
\end{figure*}

As shown below, it was inferred from the analysis of ZF spectra at 100--150 K that $\lambda$ is attributed to the background signal from muons stopped in the thermal shield (made of copper) around the sample with a small partial asymmetry $A_s\approx0.03$ (which comprises a part of $A_{\rm b}$).  Therefore, we rewrite Eq.~(\ref{Atf}) as follows:
\begin{equation}
A(t)\simeq [A_{\rm d}+(A_{\rm h}+A_{\rm s}e^{-\lambda_{\rm TF} t})]\cos\omega_\mu t,\label{Atf2}
\end{equation}
where $A_{\rm d}=A_{\rm d0}-A_{\rm b}$ refers to the contribution of the P3HT sample, and  $A_{\rm h}=A_{\rm b}-A_{\rm s}$ to that of silver sample holder.
 The curve fitting using Eq.~(\ref{Atf2}) showed excellent agreement with $A_{\rm s}$ fixed to 0.03, suggesting that the diamagnetic component coming from the sample shows negligible relaxation.  $A_{\rm d}$ increased slightly from $\sim$0.08 at 100 K to $\sim$0.10 as temperature increased [see Figs.~\ref{hfp}(a)]. In case $\lambda$ could be entirely attributed to the sample, the upper bound of the relaxation rate for the $A_{\rm d}$ component is estimated by $\lambda_{\rm d}=\lambda\cdot A_{\rm d0}/A_{\rm d}$, which  is also shown in Fig.~\ref{hfp}(b). From these evaluations, we conclude that the spin relaxation in the diamagnetic Mu contribute little to the ZF/LF-$\mu$SR spectra at least above 200 K.

In the ZF spectra at 100--150 K, the slowly relaxing $A_{\rm s}$ component is well reproduced by the dynamical Kubo-Toyabe (KT) function, $G_z^{\rm KT}(t)=G_z^\mathrm{KT}(t;\sigma,\nu_{\rm d})$ (with $\nu_{\rm d}$ being the fluctuation rate of $\sigma$), and its linewidth ($\sigma\simeq0.37$ MHz) identifies it as coming from muons stopped in a copper thermal shield \cite{Kadono:89}: the remaining non-relaxing part ($A_{\rm d}+A_{\rm h}$) is attributed to the diamagnetic Mu in P3HT and silver sample holder. We found that the KT function can be approximated by an exponential relaxation due to relatively fast fluctuations ($\nu_{\rm d}\gtrsim\sigma$). 

Here, let us summarize the separation/identification of $\mu$SR signal components: There are four components (described by the partial asymmetry), two from Mu$\dot{R}$ [$A_{\rm p}(x_{\rm p})$] and  diamagnetic Mu ($A_{\rm d}$) in the P3HT sample, and two from diamagnetic Mu (as background) in the silver sample holder ($A_{\rm h}$) and copper thermal shield ($A_{\rm s}$). Among these, $A_{\rm p}(x_{\rm p})$ and $A_{\rm s}$ respectively exhibit a characteristic field-dependence of $A_{\rm p}$ and ZF spin relaxation described by the Kubo-Toyabe function, and are thereby readily distinguished. In addition, the asymmetry of the background ($A_{\rm b}=A_{\rm h}+A_{\rm s}$) and that of the entire diamagnetic component ($A_{\rm d0}=A_{\rm d}+A_{\rm b}$) are determined by experiments, allowing the identification of the P3HT-derived diamagnetic Mu as $A_{\rm d}= A_{\rm d0}-A_{\rm b}$. Since the relaxation rate for $A_{\rm h}$ (high-purity silver) is below the lower limit of the present experiment, the only uncertainty is the distinction between the spin relaxation rate exhibited by $A_{\rm s}$ and that of $A_{\rm d}$ (particularly above $\sim$200 K), and to account for this the spin relaxation rate of the latter ($\lambda_{\rm d}$) is given as an upper bound.

The lineshape of the LF spectra suggests that the $A_{\rm p}$ component consists of multiple components with different exponential relaxation rates \cite{Risdiana:10a}, similar to those observed in NMR \cite{Yazawa:10}. Although the stretched exponential decay is often used in such situation \cite{Mashita:16}, the relaxation was approximated by a single exponential component to simplify the analysis.  A closer look at the LF spectrum shows that $A(0)$ in the spectrum at 0--10 mT is much more reduced than $A_{\rm d0}+A_{\rm p0}/2$ expected for the spin triplet Mu$\dot{R}$, where $A_{\rm p0}=A_0-A_{\rm d0}$. This indicates that the unpaired electron is subjected to NHF interactions with the surrounding protons. 
Based on these observations, the ZF/LF spectra were analyzed by least-squares fitting with the following function:
\begin{equation}
A(t)\simeq [A_{\rm p0}g_z(x_{\rm p})+A_{\rm s}]\cdot\exp[-t/T_{1\mu}]+A_{\rm h},\label{Alf}
\end{equation}
\begin{eqnarray}
g_z(x_{\rm p})&=& \frac{\frac{1}{2}h_z(x_{\rm n})+x_{\rm p}^2}{1+x_{\rm p}^2},\:\:x_{\rm p}=\gamma_{\rm av}B_{\rm LF}/\omega_0\label{gzp}\\
h_z(x_{\rm n})&\simeq& \frac{\frac{1}{3}+x_{\rm n}^2}{1+x_{\rm n}^2},\:\:x_{\rm n}\simeq\gamma_{\rm av}B_{\rm LF}/\Delta_{\rm n},
\end{eqnarray}
where $g_z(x_{\rm p})$ is the initial polarization of Mu$\dot{R}$ as a function of the normalized field $x_{\rm p}$ \cite{Patterson:88}; we presume $\omega_0\simeq2\pi\times220$ MHz and that $\omega_*$ is relatively small ($\sim2\pi\times10$ MHz) based on unpublished data \cite{Pratt:23}.  $1/T_{1\mu}$ is the relaxation rate which depends on the magnitude of $B_{\rm LF}$. 
 $h_z(x_{\rm n})$ is the approximated field dependence of initial polarization under the NHF interaction characterized by the second moment $\Delta_{\rm n}$, where the $\frac{1}{3}$ term corresponds to the possibility that the direction of the NHF field is parallel with the initial spin polarization of the triplet Mu$\dot{R}$ state \cite{Beck:75}, $\gamma_{\rm av}=(\gamma_e+\gamma_\mu)/2= 2\pi\times14.08$ MHz/mT is the gyromagnetic ratio of the unpaired electron in the triplet  state. This approximation is also supported by the fact that the spin relaxation of the triplet muonium (Mu$^0$) in solid Kr is well represented by the quasi-static KT relaxation function $G_z^{\rm KT}(t; \Delta_{\rm n},0)$ with $\gamma_\mu B_{\rm LF}$ replaced by $\gamma_{\rm av}B_{\rm LF}$, where a relatively small $\Delta_{\rm n}$ ($\approx0.7$ MHz) due to dilute $^{83}$Kr nuclei allows us to observe the detailed time evolution  \cite{Storchak:96}. In the present experimental condition with relatively large $\Delta_{\rm n}$, only the asymptotic behavior, $h_z(x_{\rm n})\approx G_z^{\rm KT}(\Delta_{\rm n}t\gg1)$, can be observed. (The exact form of $h_z(x_{\rm n})$ under LF is found elsewhere \cite{Yaouanc:10}.)  The NHF parameter in the static limit can be written more explicitly using the parameters in Eq.~(\ref{hn}) \cite{Kadono:90},
\begin{equation}
\Delta_{\rm n}^2=\sum_m\frac{1}{3}(\Omega_{\perp_m}^2+2\Omega_{\parallel_m}^2)\frac{I_m(I_m+1)}{3}.
\end{equation}
In the actual curve fits, we included the $A_{\rm s}$ component as a part of relaxing term as shown in Eq.~(\ref{Alf}), considering the difficulty to distinguish the relaxation between that exhibited by $A_{\rm s}$ and by the diamagnetic Mu in the sample. Note that $\lambda_{\rm ZF}$ in Fig.~\ref{hfp}(b) corresponds to the decay of $A_{\rm p0}g_z(0)+A_{\rm s}$. The fact that $\lambda_{\rm ZF}$ is much greater than $\lambda_{\rm TF}$ above 200 K indicates that $\lambda_{\rm ZF}$ primarily reflects spin relaxation of the Mu$\dot{R}$ state.

 As seen in Fig.~\ref{tspec}, the results of the curve fitting (with $\omega_0$ fixed to $2\pi\times 220$ MHz) reasonably reproduce the observed behavior of time spectra: the global fits for the each set of data at a given temperature, where common values of parameters ($A_{\rm p0}$, $\Delta_{\rm n}$, etc.) are assumed for the time spectra with different LFs, allowed to deduce these parameters without much uncertainty.  The LF dependence of $A(0)$ and the temperature dependence of $\Delta_{\rm n}$ obtained from these fits are shown in Figs.~\ref{hfp}(c) and \ref{hfp}(d). $\Delta_{\rm n}$ is least dependent on temperature, indicating that the radical state is stable over the entire temperature range of measurements. The mean value $\Delta_{\rm n}=2\pi\times51.9(1.4)$ MHz is consistent with the NHF parameters estimated by DFT calculations (50--55 MHz) which is dominated by the contribution of the proton situated next to Mu \cite{Pratt:23}. The model of local electronic structure based on these results is shown in Figs.~\ref{hfp}(e) and \ref{hfp}(f), where Mu is bonded to the 4-position of the thiophene ring with the neighboring proton making primary contribution to $\Delta_{\rm n}$ \cite{Pratt:23}.
 
\begin{figure}[t]
  \centering
	\includegraphics[width=0.78\linewidth,clip]{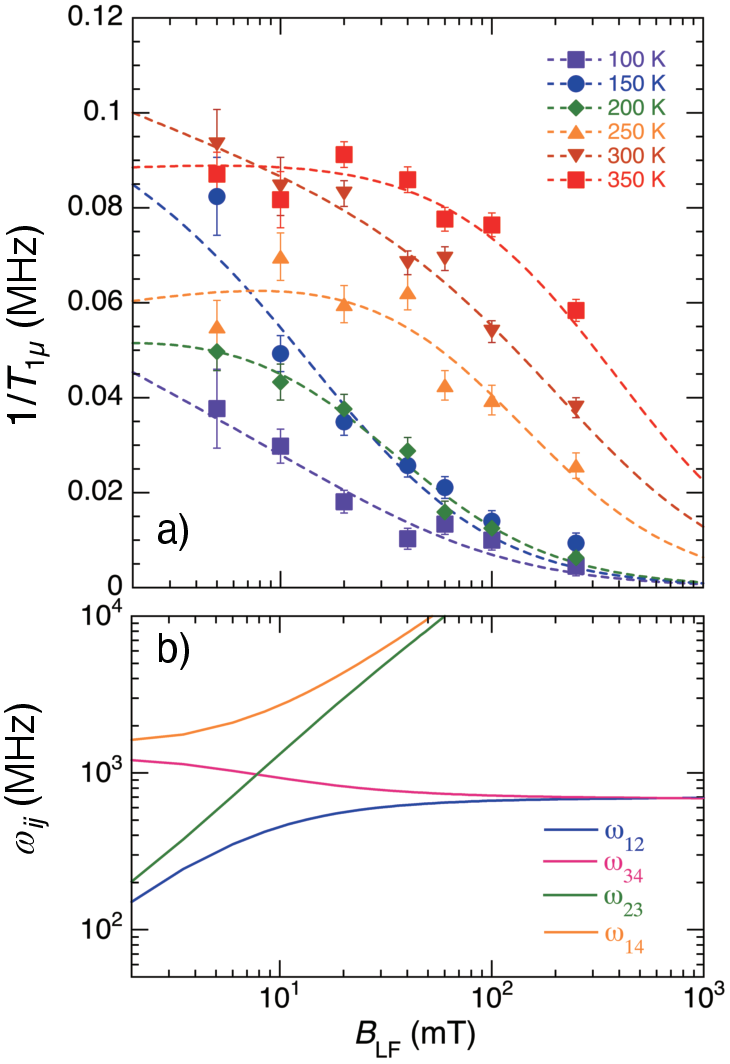}
	\caption{(a) The longitudinal relaxation rate versus $B_{\rm LF}$ exhibited by the Mu$\dot{R}$ state at various temperatures (semilog plot). The dashed lines are the results of curve fits by the Havriliak-Negami type spectral function against the magnetic field dependence (see text). (b) The angular frequencies of Zeeman transitions ($\omega_{i,j}$) for Mu$\dot{R}$ in the absence of NHF interaction ($\omega_0=2\pi\times$220 MHz). }
\label{T1}
\end{figure}

The temperature and magnetic field dependence of $1/T_{1\mu}$ obtained simultaneously in the above analysis is shown in Fig.~\ref{T1}(a) (where the ZF data were excluded for the following analysis to minimize the influence of the copper thermal shield). The dashed lines deduced from the curve fit (see below) indicate that $1/T_{1\mu}$ at 200 K and below is strongly suppressed for $B_{\rm LF}\gtrsim10^1$ mT.  Meanwhile, there is a marked increase in $1/T_{1\mu}$ on the higher magnetic field side above 200 K, which accompanies a change in the $B_{\rm LF}$ dependence from concave to convex behavior with elevated temperature.  These features seem to be difficult to describe with the conventional model for the spin dynamics. Therefore, here we first return to the basics of spin relaxation theory and introduce a model that can describe the relaxation due to more general fluctuations, and then discuss the behavior of $1/T_{1\mu}$ based on this model.

Since the spin relaxation is found to be dominated by that of Mu$\dot{R}$, it is attributed to the fluctuation of HF and/or NHF fields $H(t)$.
The relaxation rate is derived from the autocorrelation function $C(t)=\langle H(t)H(0)\rangle/\langle H(0)^2\rangle$ via the dynamical susceptibility
\begin{equation}
\chi(\omega)=\frac{1}{2}\coth\left(\frac{\hbar\omega}{2k_BT}\right)\frac{\chi_{\rm s}}{2\pi}\int C(t)e^{i\omega t}dt,
\end{equation}
\begin{equation}
\chi_{\rm s}=\Delta^2/k_BT,
\end{equation}
where the factor $\frac{1}{2}\coth(\frac{\hbar\omega}{2k_BT})$ comes from the thermal average over the canonical ensemble  (denoted by $\langle...\rangle$) \cite{Takahashi:20}, $\chi_{\rm s}$ is the static susceptibility obeying the Curie law, and $\Delta^2$ ($=\gamma_{\rm av}^2\langle H(0)^2\rangle$) is the second moment of the fluctuating local field. The general expression for the spin relaxation exhibited by Mu$\dot{R}$ is given in terms of the transitions between four energy levels,
\begin{equation} 
1/T_{1\mu} = \sum_{i,j}\frac{a_{ij}{\rm Im}\:\chi(\omega_{ij})}{\frac{1}{2}\coth(\frac{\hbar\omega_{ij}}{2k_BT})}\approx\Delta^2\sum_{i,j}\frac{a_{ij}J(\omega_{ij})}{\omega_{ij}},\label{tone}
\end{equation}
where $J(\omega)$ is the spectral density corresponding to the imaginary part of $\chi(\omega)$,  $\omega_{ij}$ are the relevant Mu$\dot{R}$ Zeeman frequencies with their respective amplitude $a_{ij}$.

In the conventional models for spin dynamics, the Lorentz-type dynamic susceptibility, $\chi(\omega)=\chi_{\rm s}/(1-i\omega/\nu)$ derived from $C(t)=\exp(-\nu t)$, is often used to yield $J(\omega)=(\omega/\nu)/[1+(\omega/\nu)^2]$, which corresponds to the Debye model in dielectric relaxation \cite{Debye:29} (or that introduced by Bloembergen-Purcell-Pound (BPP) or Redfield in NMR \cite{Bloembergen:48}). 
However, in complex systems such as polymers, the characteristic fluctuation frequencies are expected to be widely distributed. In this case, it is useful to resort to one of the most general forms that introduced by Havriliak and Negami \cite{Havriliak:67}:
\begin{equation}
\chi(\omega)=\chi_{\rm HN}(\omega)=\chi_{\rm s}\frac{1}{[1-i(\omega/\tilde{\nu})^\delta]^\gamma} 
\end{equation}
which leads to
\begin{equation}
\frac{J(\omega)}{\omega}=\frac{\sin\gamma\theta}{\omega|z|^\gamma}\simeq
\left\{
\begin{array}{ll}
\frac{\omega^{\delta-1}}{\tilde{\nu}^\delta} & (\omega\ll\tilde{\nu})\\
\frac{\tilde{\nu}^{\gamma\delta}}{\omega^{\gamma\delta+1}} &(\omega\gg\tilde{\nu})
\end{array} \label{jw}
\right.
\end{equation}
where $\gamma$ and $\delta$ are the generalized power indices ($0<\gamma,\delta\le1$), $\tilde{\nu}$ is the mean fluctuation frequency, $|z|=(x^2+y^2)^{1/2}$ with $x=1+(\omega/\tilde{\nu})^\delta\cos(\delta\pi/2)$, $y=(\omega/\tilde{\nu})^\delta\sin(\delta\pi/2)$, and $\theta=\tan^{-1}(y/x)$. An important feature of Eq.~(\ref{jw}) is that $J(\omega)/\omega$ is maximized when $\omega=\tilde{\nu}$ (i.e., the ``$T_1$ minimum'' in NMR).

Introducing $\chi_{\rm HN}(\omega)$ is also useful in considering the complexity that Mu$\dot{R}$ is subject to fluctuations of local fields at multiple Zeeman frequencies for a single magnetic field [see Fig.~\ref{T1}(b)]. For example, the minimum intra-triplet transition frequency $\omega_{12}$ is given by the following equation: 
\begin{equation}
\omega_{12} = \frac{1}{2}\omega_0\left[1-(1+x_{\rm p}^2)^{1/2}\right]+\omega_-,
\end{equation}
where $\omega_-=\gamma_{\rm av}B_{\rm LF}$.
Therefore, $\omega_{12}\simeq\omega_-$ for $x_{\rm p}\lesssim1$ ($B_{\rm LF}\lesssim18$ mT) and $\omega_{12}\simeq\omega_0/2-\omega_\mu$ for $x_{\rm p}\gg1$ ($B_{\rm LF}\gg 18$ mT), indicating that $\omega_{12}$ changes nonlinearly with respect to $B_{\rm LF}$.  Considering the foreseeable complexity for the behavior of $\tilde{\nu}$ in polymers, it is not realistic to compare $1/T_{1\mu}$ calculated using Eq.~(\ref{tone}) with experimental values. Therefore, we treat the obtained $1/T_{1\mu}$ as a function of the magnetic field, namely
\begin{equation}
1/T_{1\mu}\simeq\Delta^2\frac{J(\omega)}{\gamma_{\rm av}B_{\rm LF}}, \:\:\:\:\omega/\tilde{\nu}=B_{\rm LF}/B_{\tilde{\nu}},\label{toneb}
\end{equation}
and attempt to characterize it in terms of the parameters in the Havriliak-Negami function; here, $B_{\tilde{\nu}}$ is the characteristic field defined by $\tilde{\nu}=\gamma_{\rm av}B_{\tilde{\nu}}$. We then discuss the nature of the fluctuations while semi-quantitatively considering the relationship with the Zeeman level of Mu$\dot{R}$.  

\begin{figure}[t]
  \centering
	\includegraphics[width=0.75\linewidth,clip]{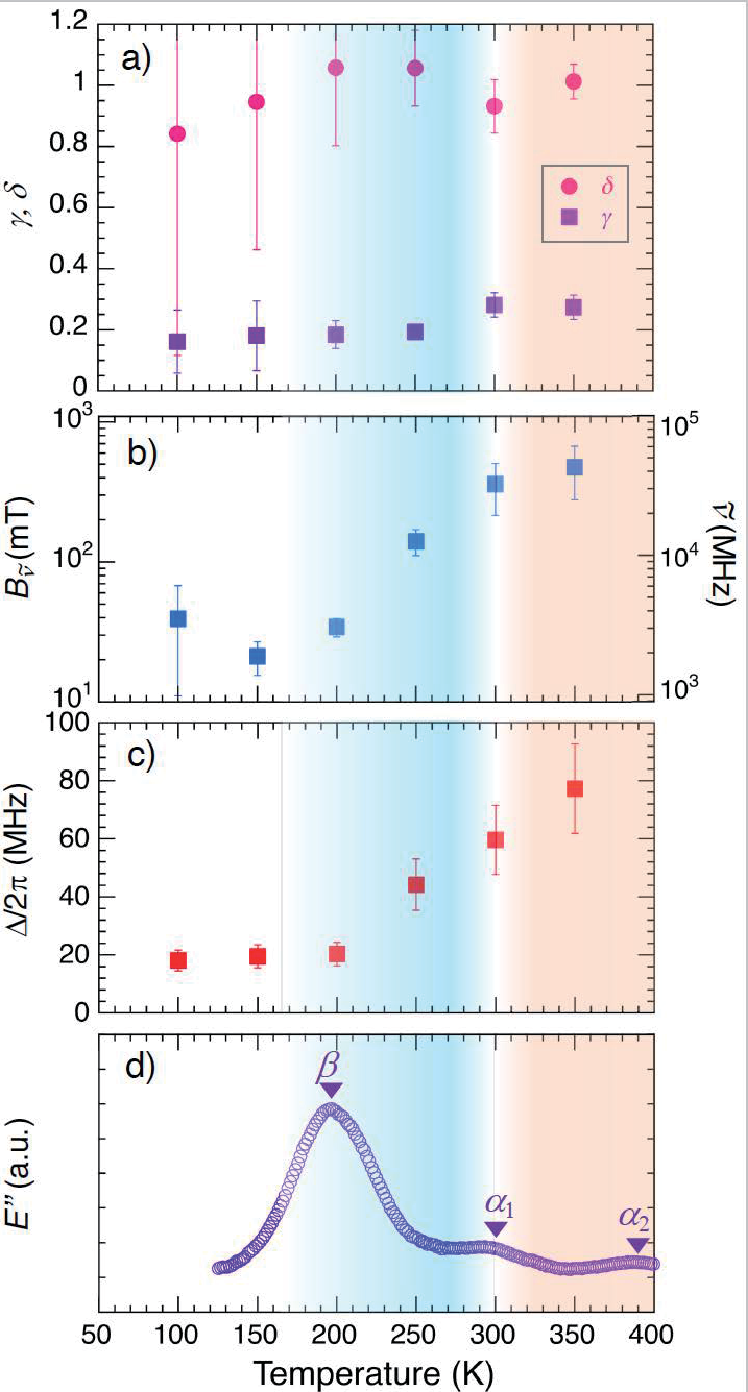}
	\caption{ The parameters in the Havriliak-Negami function obtained by curve fits shown in Fig.~\ref{T1}: (a) power indices, (b) mean field (with corresponding mean frequency in the right axis), and (c) hyperfine parameter. (d) Dynamic loss modulus ($E''$) for a P3HT film at a frequency of 20 Hz, where three relaxation peaks corresponding to $\beta$, $\alpha_1$, and $\alpha_2$ processes are observed at 200, 300 and 390 K \cite{Ogata:15}. The colored bands (200--300 K and above $\sim$300 K) indicate the temperature regions discerned by the difference in the mode of molecular motion inferred from $^{13}$C-NMR \cite{Yazawa:10} and $E''$.}
\label{abnu}
\end{figure}

The dashed lines in Fig.~\ref{T1} show the results of curve fitting by the least-squares method assuming Eqs.~(\ref{jw}) and (\ref{toneb}). These curves reasonably reproduce the magnetic field dependence of the data, although the corresponding parameters exhibit considerable uncertainty due to the relatively strong mutual correlations. The temperature dependence of $\gamma$, $\delta$, $B_{\tilde{\nu}}$, and $\Delta$ obtained from this analysis is shown in Fig.~\ref{abnu}.

According to previous $^{13}$C-NMR and FT-IR studies, the behavior of the P3HT chains can be roughly divided into three regions in the temperature range observed here. At low temperatures below 200 K, both the thiophene ring and the hexyl side chain are in a frozen state, while at 200--300 K, only the hexyl side chain is in motion (crystalline state). It is further suggested that above 300 K, in addition to the motion of the hexyl side chain, the twisting motion of the thiophene ring is also excited (plastic crystal state). It is noteworthy that most of the parameters in Fig.~\ref{abnu} appear to exhibit correlations with such a change in the motional mode: $B_{\tilde{\nu}}$  and $\Delta$ gradually increases above 200 K; $\delta$ and $\gamma$ respectively exhibit a slight increase above 200 K and 300 K (although the behavior is blurred by large error bars reflecting uncertainty in the curve fits). These correlations suggest that the spin relaxation seen in $\mu$SR originates from molecular motion. Note that the values of $\gamma$ and $\delta$ are similar to those of other polymers obtained from dielectric relaxation analysis at the same ambient temperature \cite{Havriliak:67}.

Let us now examine these changes more quantitatively: in Eq.~(\ref{jw}), the relatively small $\gamma$ and $\delta$ below $\sim$200 K indicate that $J(\omega)$ spreads over the wide region of $\omega$ centered at $\tilde{\nu}$. Thus, they imply a slightly broader spectral density distribution at low temperatures. However, above 200 K, $\delta\simeq1$ and $B_{\tilde{\nu}}$ shows gradual increase from $\sim$10$^1$ to $10^2$ mT. This suggests that fluctuations with frequencies around $\tilde{\nu}$ (see right vertical axis of Fig.~\ref{abnu}(b)) become dominant in $J(\omega)$ in this temperature region. 

It is reported in the previous $^{13}$C-NMR study that the temperature dependence of the longitudinal relaxation rate ($1/T_1$) for the $^{13}$C signals in the hexyl side chain exhibit broad peaks centered around 300 K (i.e., the $T_1$ minimum), which indicates that the fluctuation rate around this temperature is equal to the Zeeman angular frequency $\omega_{\rm NMR}\approx10^3$ MHz \cite{Yazawa:10}. This is in reasonable agreement with $\tilde{\nu}\simeq10^3$--$10^4$ MHz at 200--300 K, and suggesting that $1/T_{1\mu}$ is also dominated by hexyl chain motion.  Above $\sim$300 K, $B_{\tilde{\nu}}$ reaches 3--$4\times10^2$ mT ($\tilde{\nu}\gtrsim10^4$ MHz) with $\Delta$ exceeding $\Delta_{\rm n}$, suggesting that the contribution of thiophene ring motion sets in. The $^{13}$C-NMR results also suggest that $1/T_1$ of $^{13}$C signals in the thiophene ring continues to increase with increasing temperature over the range from 170 K to 370 K, where the fluctuation rate of the thiophene ring increases on a scale 1--2 orders of magnitude smaller than that of the hexyl side chain. This indicates that the contribution of the thiophene ring motion to $1/T_{1\mu}$ is expected to be observed only at higher temperatures above $\sim$300 K, which is also consistent with the present experimental results and their interpretation.

 \section{DISCUSSION}\label{Dcn}

By introducing the Havriliak-Negami function into the spectral density function $J(\omega)$ describing $1/T_{1\mu}$, we were able to identify molecular motion as the microscopic origin of $1/T_{1\mu}$ on the semi-quantitative basis. In the following, we will examine the consistency of parameters in the dynamical model with respect to the mean fluctuation frequency $\tilde{\nu}$.

As shown in Fig~\ref{abnu}(c), $\Delta$ is considerably smaller than $\Delta_{\rm n}$ below 200 K ($\Delta\approx 2\pi\times20$ MHz), and gradually increases with increasing temperature.  This is in contrast to the naive expectation that $\Delta$ is  independent of temperature and equal to $\Delta_{\rm n}$. However, as can be inferred from Figs.~\ref{hfp}(e) and (f), the motion of molecules in the crystal is spatially restricted, so that the fluctuations in $H(t)$ induced by the associated relative motion of protons and unpaired electrons are also partially suppressed.  Under such circumstances, a part of the autocorrelation of $H(t)$ is non-vanishing for $t\rightarrow\infty$, and the observed $\Delta$ may correspond to the magnitude of the remaining fluctuating component.  
More specifically, in the case of Lorentz-type $\chi(\omega)$, this situation can be described by the Edwards--Anderson order parameter $Q$ in the autocorrelation function \cite{Edwards:75,Edwards:76},
\begin{equation}
\frac{\langle H(t)H(0)\rangle}{\langle H(0)^2\rangle} = \frac{\langle H(t)H(0)\rangle}{\Delta_{\rm n}^2/\gamma_{\rm av}^2} = Qe^{-\nu t}+(1-Q).\label{EAp}
\end{equation}
The fact that $\Delta$ is smaller than $\Delta_{\rm n}$ is consistent with this situation, $\Delta=\sqrt{Q}\Delta_{\rm n}$, where $\sqrt{Q}=\Delta/\Delta_{\rm n}\approx0.4$ (i.e., $Q\approx0.16$) below 200 K. 
It is then interpreted that $\Delta$ below $\sim$200 K mainly reflects relatively restricted fluctuations of protons on the hexyl chain which becomes progressively more intense as the temperature rises above $\sim$200 K. 

On the other hand, it is also possible that $\Delta$ is dominated by the fluctuation of $\omega_0$ (and/or $\omega_*$), as suggested for isolated molecules in the gas phase or in solution \cite{Freming:96,McKenzie:11}.  In this case, $\Delta$ is expected to be as large as $\omega_0$ for free rotation/vibration of molecules which is far greater than the observed $\Delta$. However, provided that only a tiny part of it fluctuates for subtle twist motions, etc., we can use the same scenario as for the NHF field above, where $\Delta\approx \sqrt{Q}\omega_0$.  Assuming an even smaller value of $Q$ than in the NHF case [$\approx0.01$ for the Lorentz-type $\chi$] is sufficient to explain the observed magnitude of $\Delta$. The trend seen in Fig.~\ref{abnu}(c) of $\Delta$ increasing over $\Delta_{\rm n}$ above 300 K may suggest an additional contribution from the fluctuating HF field.

Now, the above interpretation gives rise to another issue regarding the behavior of the initial asymmetry $A_{\rm p}$ vs.~$B_{\rm LF}$. As can be seen from Figs.~\ref{abnu}(b) and (c), the mean frequency $\tilde{\nu}$ is larger than $\Delta$ at all temperatures, indicating that spin relaxation is subject to the motional narrowing effect. This appears at first glance to contradict the behavior of $A_{\rm p}$ shown in Fig.~\ref{hfp}(c) (which is common over the entire range of temperature, as is clear from Fig.~\ref{tspec}).   Namely, while $A_{\rm p}$ at zero field would decrease to $A_{\rm p0}g_z(0)\approx A_{\rm p0}/6$ for $\tilde{\nu}\lesssim\Delta$ due to relaxation by the static distribution of $H(t)$, the effective $\Delta$ is expected to decrease due to so-called motional narrowing for $\tilde{\nu}>\Delta$, and $A_{\rm p}$ at zero field is expected to recover to $A_{\rm p0}/2$.  

\begin{figure}[t]
  \centering
	\includegraphics[width=0.95\linewidth,clip]{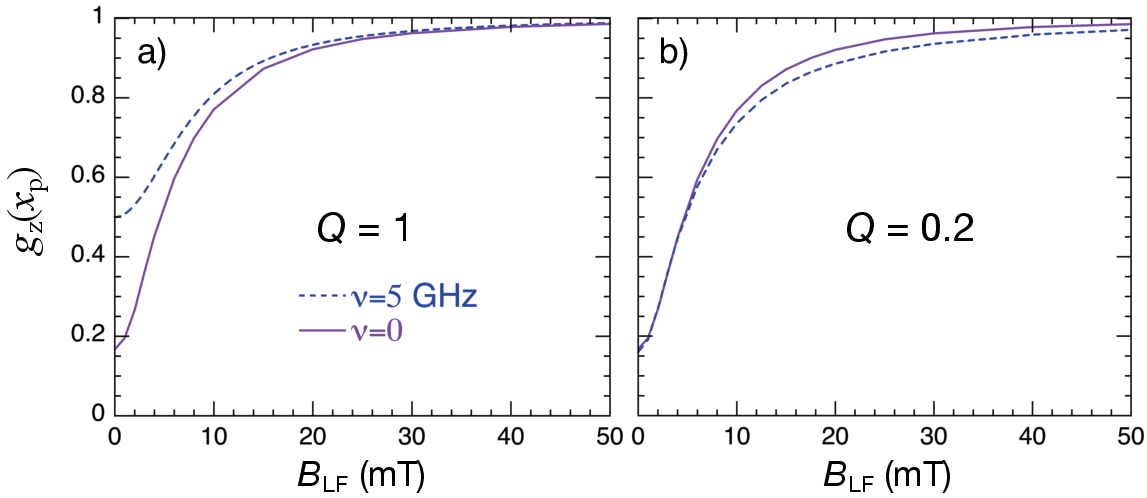}
	\caption{
	Simulation of the initial polarization $g_z(x_{\rm p})$ for Mu$\dot{R}$, where $\omega_0=2\pi\times220$ MHz and $\Delta_{\rm n}=2\pi\times50$ MHz.  The longitudinal field dependence of $g_z(x_{\rm p})$ corresponding to fluctuations $\nu=0$ and 5 GHz is shown for $Q=1$  (a) and 0.2 (b). 
	}
\label{LFsim}
\end{figure}

However, such a quasistatic behavior of $A_{\rm p}$ is expected when the dynamical and static components coexist in the fluctuations of $H(t)$, corresponding to $Q<1$ in Eq.~(\ref{EAp}). We extended the recently proposed dynamical model for diamagnetic Mu based on Eq.~(\ref{EAp}) \cite{Ito:24} to the present case (where $\tilde{\nu}$ is replaced with $\nu$) and the simulation results are shown in Fig.~\ref{LFsim}: when the full amplitude $\Delta_{\rm n}$ fluctuates ($Q=1$), the initial polarization $g_z(x_{\rm p})$ converges to $\frac{1}{2}$ at $x_{\rm p}=0$ for $\nu>\Delta_{\rm n}$, indicating that the static depolarization due to $\Delta_{\rm n}$ is quenched by the motional narrowing. On the other hand, for $Q=0.2$, the field dependence of $g_z(x_{\rm p})$ is close to that for the quasistatic muonium and almost independent of $\nu$.
Similarly, when $\omega_0$ in Mu$\dot{R}$ fluctuates in its full magnitude (i.e., $Q=1$), it is predicted that $g_z(x_{\rm p})\rightarrow 1$ for $\nu\gg\omega_0$, i.e., Mu$\dot{R}$  becomes effectively equivalent with the diamagnetic Mu in terms of magnetic response. (This situation cannot be discerned by the behavior of $1/T_{1\mu}$ alone, so that special care must be taken for that of the initial asymmetry.)

The same situation has already been demonstrated in the simulation for the diamagnetic Mu which exhibits depolarization due to random local fields from surrounding nuclear magnetic moments. The behavior of relaxation function with respect to $\nu$ deviates from that of the dynamical KT function for $Q<1$, and it becomes less dependent on $\nu$ and converges to a quasi-static KT function as $Q$ approaches zero. An important point to note in this simulation is that $Q=1$ is expected when the fluctuations are due to self-diffusion of the diamagnetic Mu, while $Q<1$ when they are governed by the dynamics of the surrounding ions \cite{Ito:24}.  Since the similar situation can be presumed for the paramagnetic Mu, the small $Q$ suggested in P3HT can be regarded as evidence that the cause of the fluctuations is not the hopping motion of Mu$\dot{R}$ itself.

In view of the above discussion, it is necessary to reconsider the interpretation claimed in previous studies that the muon $1/T_{1\mu}$ observed in P3HT is due to one-dimensional (1D) motion of unpaired electron on the conjugated molecular chain \cite{Risdiana:10a,Risdiana:10b}.  The main rationale in these studies is that the $B_{\rm LF}$ dependence of $1/T_{1\mu}$ at low temperatures follows $J(\omega)/\omega\propto\omega^{-1/2}$ as theoretically predicted for 1D jumping motion of electrons \cite{Risch:92}. However, the linewidth $\Delta$ for the corresponding fluctuation is expected to be comparable to $\omega_0$ with large $Q$ ($\simeq1$).  Moreover, the 1D jump frequency is reported to be $\nu\simeq5.3\times10^{14}$ s$^{-1}$ from NMR \cite{Mabboux:95}, which is much greater than $\omega_0$ ($\simeq1.38\times10^9$ s$^{-1}$).  Therefore, if the unpaired electron associated with Mu$\dot{R}$ are moving at such high jumping frequencies, it is expected that $A_{\rm p}\simeq A_{\rm p0}$ [i.e., $g_z(x_{\rm p})\simeq1$] regardless of $B_{\rm LF}$; see the dashed  line in Fig.~\ref{hfp}(c). A similar behavior of $g_z(x_{\rm p})$ has also been shown theoretically \cite{Nosov:63,Risch:92,Patterson:88}, which is inconsistent with the observations that  $A_{\rm p}$ strongly depends on $B_{\rm LF}$. Another issue is that they report the large deviation of $1/T_{1\mu}$ from this $\omega^{-1/2}$ behavior around room temperature, which they attribute to the onset of three-dimensional (3D) diffusion of carriers \cite{Risdiana:10a,Risdiana:10b}. However, this is not consistent with the implication from NMR that the 1D-3D anisotropy remains large ($D_\parallel/D_\perp\sim10^6$) in the relevant temperature range \cite{Mabboux:95}. Needless to mention, there is no such contradiction in our interpretation that $T_{1\mu}$ is due to molecular motion rather than unpaired electron motion.

\section{Summary \& Conclusion}
We have shown that muons implanted to crystalline P3HT fall into a muoniated radical state and a diamagnetic state, and that the unpaired electron in the former undergoes HF interaction with muons bound to the thiophene ring and NHF interaction with surrounding protons.  Furthermore, we have succeeded in observing the local molecular motions of P3HT from longitudinal spin relaxation due to fluctuations of the HF and/or NHF fields. By introducing the Havriliak-Negami function for the interpretation of the spectral density $J(\omega)$ inferred from the LF-$\mu$SR measurements, we have been able to make a semi-quantitative correspondence between the phenomenology of the spectral density in dielectric relaxation and the microscopic origin of muon spin relaxation due to molecular motion. This also served to establish a basis for comparison of $\mu$SR result with those of other microscopic probes such as NMR. The detailed magnetic field dependence measurements of LF-$\mu$SR have revealed a qualitative change in $J(\omega)$ with temperature, which is consistent with the $^{13}$C-NMR result that revealed the hexyl chain motion at 200-300 K and the twisting motion of the thiophene ring above 300 K. Similar studies are expected to be widely applicable to other macromolecular systems.

\begin{acknowledgments}
This work was supported by the Photon and Quantum Basic Research Coordinated Development (QBRCD) Program from the Ministry of Education, Culture, Sports, Science, and Technology (MEXT), and partly by the JST-Mirai Program (JPMJMI18A2), Japan. The authors would like to thank F. Pratt for providing information on $\mu$SR experiment and DFT calculations for muonated radicals in P3HT prior to publication. Thanks are also to H. Seto for acquisition of the financial support from the QBRCD Program, and to the MUSE staff for the help during experiment at MLF, J-PARC. The muon experiment was conducted under the support of Inter-University-Research Programs by Institute of Materials Structure Science, KEK (Proposal No.~2014MS03).
\end{acknowledgments}
\section*{References}
\vspace{-0.25in}
%
\end{document}